\documentclass[twocolumn,preprintnumbers]{revtex4-1}
\pdfoutput=1

\usepackage{amsmath,amssymb,graphicx,booktabs,bm,psfrag,color}

\begin{document}

\title{\boldmath One Leptoquark to Rule Them All:\\
A Minimal Explanation for $R_{D^{(*)}}$, $R_K$ and $(g-2)_\mu$}

\preprint{MITP/15-100}
\preprint{November 9, 2015}

\author{Martin Bauer$^a$}
\author{Matthias Neubert$^{b,c}$}

\affiliation{$^a$Institut f\"ur Theoretische Physik, Universit\"at Heidelberg, Philosophenweg 16, 69120 Heidelberg, Germany\\
${}^b$PRISMA Cluster of Excellence {\em\&} MITP, Johannes Gutenberg University, 55099 Mainz, Germany\\
${}^c$Department of Physics {\em\&} LEPP, Cornell University, Ithaca, NY 14853, U.S.A.}

\begin{abstract}
We show that by adding a single new scalar particle to the Standard Model, a TeV-scale leptoquark with the quantum numbers of a right-handed down quark, one can explain in a natural way three of the most striking anomalies of particle physics: the violation of lepton universality in $\bar B\to\bar K\ell^+\ell^-$ decays, the enhanced $\bar B\to D^{(*)}\tau\bar\nu$ decay rates, and the anomalous magnetic moment of the muon. Constraints from other precision measurements in the flavor sector can be satisfied without fine-tuning. Our model predicts enhanced $\bar B\to\bar K^{(*)}\nu\bar\nu$ decay rates and a new-physics contribution to $B_s\!-\!\bar B_s$ mixing close to the current central fit value.
\end{abstract}

\maketitle

{\em Introduction.} 
Rare decays and low-energy precision measurements provide powerful probes of physics beyond the Standard Model (SM). During the first run of the LHC, many existing measurements of such observables were improved and new channels were discovered, at rates largely consistent with SM predictions. However, a few anomalies observed by previous experiments have been reinforced by LHC measurements and some new anomalous signals have been reported. The most remarkable example of a confirmed effect is the $3.5\sigma$ deviation from the SM expectation in the combination of the ratios
\begin{equation}
   R_{D^{(*)}} 
   = \frac{\Gamma(\bar B\to D^{(*)}\tau\bar\nu)}{\Gamma(\bar B\to D^{(*)}\ell\bar\nu)} \,; \quad
   \ell=e,\mu.
\end{equation}
An excess of the $\bar B\to D^{(*)}\tau\bar\nu$ decay rates was first noted by BaBar \cite{Lees:2012xj,Lees:2013uzd}, and it was shown that this effect cannot be explained in terms of type-II two Higgs-doublet models. The relevant rate measurements were consistent with those reported by Belle \cite{Matyja:2007kt,Adachi:2009qg,Bozek:2010xy} and were recently confirmed by LHC$b$ for the case of $R_{D^*}$ \cite{Aaij:2015yra}. Since these decays are mediated at tree level in the SM, relatively large new-physics contributions are necessary in order to explain the deviations. Taking into account the differential distributions $d\Gamma(\bar B\to D\tau\bar\nu)/dq^2$ provided by BaBar \cite{Lees:2013uzd} and Belle \cite{Huschle:2015rga}, only very few models can explain the excess, and they typically require new particles with masses near the TeV scale and $O(1)$ couplings \cite{Fajfer:2012jt,Crivellin:2012ye,Celis:2012dk,Tanaka:2012nw,Dorsner:2013tla,Sakaki:2013bfa,Bhattacharya:2014wla,Alonso:2015sja,Calibbi:2015kma,Freytsis:2015qca}. One of the interesting new anomalies is the striking $2.6\sigma$ departure from lepton universality of the ratio
\begin{equation}\label{RK}
   R_K = \frac{\Gamma(\bar B\to\bar K\mu^+\mu^-)}{\Gamma(\bar B\to\bar K e^+ e^-)}
   = 0.745\,_{-0.074}^{+0.090}\pm 0.036
\end{equation}
in the dilepton invariant mass bin $1\,{\rm GeV}^2\le q^2\le 6\,{\rm GeV}^2$, reported by LHC$b$ \cite{Aaij:2014ora}. This ratio is essentially free from hadronic uncertainties, making it very sensitive to new physics. Equally intriguing is a discrepancy in angular observables in the rare decays $\bar B\to\bar K^*\mu^+\mu^-$ seen by LHC$b$ \cite{Aaij:2013qta}, which is however subject to significant hadronic uncertainties \cite{Beaujean:2013soa, Lyon:2014hpa, Jager:2014rwa}. Both observables are induced by loop-mediated processes in the SM, and assuming $O(1)$ couplings one finds that the dimension-6 operators that improve the global fit to the data are suppressed by mass scales of order tens of TeV \cite{Hurth:2014vma,Altmannshofer:2014rta,Beaujean:2015gba,Descotes-Genon:2015uva}.

In this letter we propose a simple extension of the SM by a single scalar leptoquark $\phi$ transforming as $(\bm{3},\bm{1},-\frac13)$ under the SM gauge group, which can explain both the $R_{D^{(*)}}$ and the $R_K$ anomalies with a low mass $M_\phi\sim 1$~TeV and $O(1)$ couplings. The fact that such a particle can explain the anomalous $\bar B\to D^{(*)}\tau\bar\nu$ rates and $q^2$ distributions is well known \cite{Sakaki:2013bfa,Freytsis:2015qca}. Here we show that the same leptoquark can resolve in a natural way the $R_K$ anomaly and explain the anomalous magnetic moment of the muon. Reproducing $R_K$ with a light leptoquark is possible in our model, because the transitions $b\to s\ell^+\ell^-$ are only mediated at loop level. Such loop effects have not been studied previously in the literature. We also discuss possible contributions to $B_s\!-\!\bar B_s$ mixing, the rare decays $\bar B\to\bar K^{(*)}\nu\bar\nu$, $D^0\to\mu^+\mu^-$, $\tau\to\mu\gamma$, and the $Z$-boson couplings to fermions. We focus primarily on fermions of the second and third generations, leaving a more complete analysis for future work.

The leptoquark $\phi$ can couple to $LQ$ and $e_R u_R$, as well as to operators which would allow for proton decay and will be ignored in the following. Such operators can be eliminated, e.g., by means of a discrete symmetry, under which SM leptons and $\phi$ are assigned opposite parity. The leptoquark interactions follow from the Lagrangian
\begin{equation}
\begin{aligned}
   {\cal L}_\phi &= (D_\mu\phi)^\dagger D_\mu\phi - M_\phi^2\,|\phi|^2 - g_{h\phi}\,|\Phi|^2 |\phi|^2 \\
   &\quad\mbox{}+ \bar Q^c\bm{\lambda}^L i\tau_2 L\,\phi^*
    + \bar u_R^c\,\bm{\lambda}^R e_R\,\phi^* + \mbox{h.c.} \,,
\end{aligned}
\end{equation}
where $\Phi$ is the Higgs doublet, $\bm{\lambda}^{L,R}$ are matrices in flavor space, and $\psi^c=C\bar\psi^T$ are charge-conjugate spinors. Note that our leptoquark shares the quantum numbers of a right-handed sbottom, and the couplings proportional to $\bm{\lambda}^L$ can be reproduced from the R-parity violating superpotential. The above Lagrangian refers to the weak basis. Switching to the mass basis for quarks and charged leptons, the couplings to fermions take the form
\begin{equation}\label{Leff}
   {\cal L}_\phi \ni \bar u_L^c\bm{\lambda}_{ue}^L e_L\,\phi^* 
    - \bar d_L^c\bm{\lambda}_{d\nu}^L\nu_L \phi^*
    + \bar u_R^c\,\bm{\lambda}_{ue}^R e_R\,\phi^* + \mbox{h.c.} \,,
\end{equation}
where
\begin{equation}
   \bm{\lambda}_{ue}^L = \bm{U}_u^T \bm{\lambda}^L \bm{U}_e \,, ~~
   \bm{\lambda}_{d\nu}^L = \bm{U}_d^T \bm{\lambda}^L \,, ~~
   \bm{\lambda}_{ue}^R = \bm{V}_u^T \bm{\lambda}_R \bm{V}_e \,,
\end{equation}
and $\bm{U}_q$ ($\bm{V}_q$) denote the rotations of the left-handed (right-handed) fermion fields. These definitions imply
\begin{equation}\label{cute}
   \bm{V}_{\rm CKM}^T\,\bm{\lambda}_{ue}^L = \bm{\lambda}_{d\nu}^L \bm{U}_{e} \,,
\end{equation}
which involves the CKM matrix $\bm{V}_{\rm CKM}=\bm{U}_u^\dagger \bm{U}_d$. ATLAS and CMS have searched for pair-produced leptoquarks in various final states. The search channels $\phi\phi^*\to\mu^+\mu^- jj$ and $\phi\phi^*\to b\bar b\nu\bar\nu$ are the most relevant ones for our analysis. The most recent ATLAS/CMS analyses exclude a leptoquark lighter than 850~GeV/760~GeV at 95\% CL, assuming $\mbox{Br}(\phi\to\mu j)=0.5$ \cite{Aad:2015caa,Khachatryan:2015vaa}. ATLAS also derives a lower bound of 625~GeV assuming $\mbox{Br}(\phi\to b\nu)=1$ \cite{Aad:2015caa}. These bounds can be weakened by reducing the branching fractions to the relevant final states.

\begin{figure}
\includegraphics[width=0.4\textwidth]{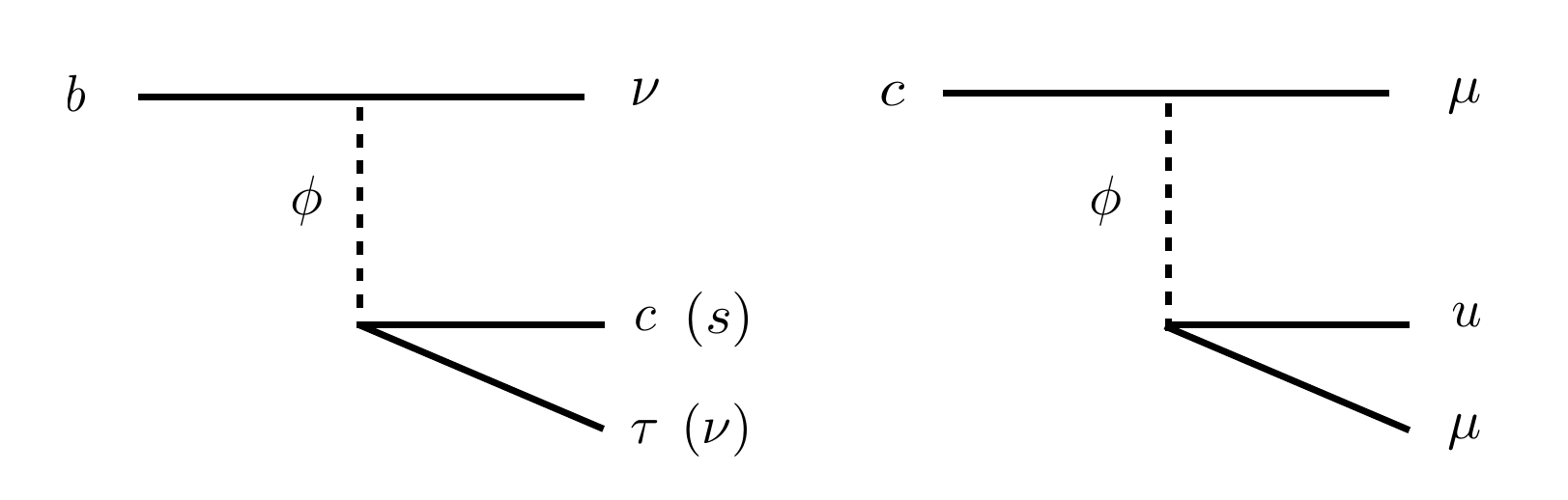}
\vspace{-1mm}
\caption{\label{fig:treegraphs}
Tree-level diagrams contributing to weak decays.}
\end{figure}

\vspace{2mm}
{\em Tree-Level Processes.} 
The leptoquark $\phi$ mediates semileptonic $B$-meson decays at tree level, as shown in the first graph of Figure~\ref{fig:treegraphs}. This gives rise to the effective Lagrangian 
\begin{align}\label{eq:sign}
   {\cal L}_{\rm eff}^{(\phi)}
   &= \frac{1}{2M_\phi^2} \bigg[ 
    - \lambda_{u_i\ell_j}^{L*}\lambda_{b\nu_k}^L 
    \bar u_L^i\gamma_\mu b_L\,\bar\ell_L^j\gamma^\mu\nu_L^k \\
   &\mbox{}+ \lambda_{u_i\ell_j}^{R*}\lambda_{b\nu_k}^L 
    \bigg( \bar u_R^i b_L\,\bar\ell_R^j \nu_L^k
    - \frac{\bar u_R^i\sigma_{\mu\nu} b_L\,\bar\ell_R^j\sigma^{\mu\nu} \nu_L^k}{4} \bigg) \bigg] \,, \notag
\end{align}
where $i,j,k$ are flavor indices. The first term generates additive contributions to the CKM matrix elements $V_{ub}$ and $V_{cb}$, which may be different for the different lepton flavors. The second term includes novel tensor structures not present in the SM. It may help to explain why determinations of $V_{ub}$ and $V_{cb}$ from inclusive and exclusive $B$-meson decays give rise to different results. Of particular interest are the decays $\bar B\to D^{(*)}\tau\bar\nu$, whose rates are found to be about 30\% larger than in the SM. A model-independent analysis of this anomaly in the context of effective operators, including the effects of renormalization-group (RG) evolution from $\mu=M_\phi$ to $\mu=m_b$, has been performed in \cite{Sakaki:2013bfa,Freytsis:2015qca}. In the last paper it was found that an excellent fit to the experimental data is obtained for a scalar leptoquark with parameters 
\begin{equation}\label{RDsolu}
   \lambda_{c\tau}^{L*}\lambda_{b\nu_\tau}^L\approx 0.35\,\hat M_\phi^2 \,, \quad
   \lambda_{c\tau}^{R*}\lambda_{b\nu_\tau}^L\approx -0.03\,\hat M_\phi^2
\end{equation}
with large and anti-correlated errors, where it was assumed that the only relevant neutrino is $\nu_\tau$, as only this amplitude can interfere with the SM and hence give rise to a large effect. Throughout this letter $\hat M_\phi\equiv M_\phi/{\rm TeV}$. For a leptoquark mass near the TeV scale, these conditions can naturally be satisfied with $O(1)$ left-handed and somewhat smaller right-handed couplings. We will ignore the three other fit solutions found in \cite{Freytsis:2015qca}, since they require significantly larger couplings. 

Our model also gives rise to tree-level flavor-changing neutral currents (FCNCs), some examples of which are shown in Figure~\ref{fig:treegraphs}. Particularly important for our analysis are the rare decays $\bar B\to\bar K\nu\bar\nu$ and $D^0\to\mu^+\mu^-$. The effective Lagrangian for $\bar B\to\bar K^{(*)}\nu\bar\nu$ as well as the corresponding inclusive decay reads
\begin{equation}
   {\cal L}_{\rm eff}^{(\phi)}
   = \frac{1}{2M_\phi^2}\,\lambda_{s\nu_i}^{L*}\lambda_{b\nu_j}^L\, 
    \bar s_L\gamma_\mu b_L\,\bar\nu_L^i\gamma^\mu\nu_L^j \,.
\end{equation}
Apart from possibly different neutrino flavors, this involves the same operator as in the SM. It follows that the ratio $R_{\nu\bar\nu}=\Gamma/\Gamma_{\rm SM}$ for either the exclusive or the inclusive decays is given by 
\begin{equation}\label{Rvv}
   R_{\nu\bar\nu}^{(\phi)} 
   = 1 - \frac{2r}{3}\,\mbox{Re}\,\frac{\big(\lambda^L\lambda^{L\dagger}\big)_{bs}}{V_{tb} V_{ts}^*}
    + \frac{r^2}{3}\,
    \frac{\big(\lambda^L\lambda^{L\dagger}\big)_{bb} \big(\lambda^L\lambda^{L\dagger}\big)_{ss}}%
         {\big| V_{tb} V_{ts}^* \big|^2} \,,
\end{equation}
where $\big(\lambda^L\lambda^{L\dagger}\big)_{bs}=\sum_i\lambda_{b\nu_i}^L\lambda_{s\nu_i}^{L*}$ etc., and
\begin{equation}
   r = \frac{s_W^4}{2\alpha^2}\,\frac{1}{X_0(x_t)}\,\frac{m_W^2}{M_\phi^2}
   \approx \frac{1.91}{\hat M_\phi^2} \,.
\end{equation}
Here $X_0(x_t)=\frac{x_t(2+x_t)}{8(x_t-1)}+\frac{3x_t(x_t-2)}{8(1-x_t)^2}\ln x_t\approx 1.48$ with $x_t=m_t^2/m_W^2$ denotes the SM loop function, and $s_W^2=0.2313$ is the sine squared of the weak mixing angle. Currently the strongest constraint arises from upper bounds on the exclusive modes $B^-\to K^-\nu\bar\nu$ and $B^-\to K^{*-}\nu\bar\nu$ obtained by BaBar \cite{Lees:2013kla} and Belle \cite{Lutz:2013ftz}, which yield $R_{\nu\bar\nu}<4.3$ and $R_{\nu\bar\nu}<4.4$ at 90\% CL \cite{Buras:2014fpa}. Using the Schwarz inequality, we then obtain from (\ref{Rvv}) 
\begin{equation}\label{bound1}
   - 1.20\,\hat M_\phi^2 < \mbox{Re}\,\frac{\big(\lambda^L\lambda^{L\dagger}\big)_{bs}}{V_{tb} V_{ts}^*}
   < 2.25\,\hat M_\phi^2 \,.
\end{equation}
The FCNC process $D^0\to\mu^+\mu^-$ can arise at tree level in our model. Neglecting the SM contribution, which is two orders of magnitude smaller than the current experimental upper bound, we find the decay rate
\begin{align}
   \Gamma &= \frac{f_D^2\,m_D^3}{256\pi M_\phi^4} \left( \frac{m_D}{m_c} \right)^2 \!\beta_\mu
    \Bigg[ \beta_\mu^2\,\big| \lambda_{c\mu}^L \lambda_{u\mu}^{R*} - \lambda_{c\mu}^R \lambda_{u\mu}^{L*} \big|^2 \\
   &\mbox{}+ \bigg| \lambda_{c\mu}^L \lambda_{u\mu}^{R*} \!+\! \lambda_{c\mu}^R \lambda_{u\mu}^{L*}
    + \frac{2m_\mu m_c}{m_D^2} \big(
    \lambda_{c\mu}^L \lambda_{u\mu}^{L*} \!+\! \lambda_{c\mu}^R \lambda_{u\mu}^{R*} \big) \bigg|^2 \Bigg] , \notag
\end{align}
where $f_D=212(1)$~MeV \cite{Rosner:2015wva} is the $D$-meson decay constant and $\beta_\mu=(1-4m_\mu^2/m_D^2)^{1/2}$. We use the running charm-quark mass $m_c\equiv m_c(M_\phi)\approx 0.54$~GeV to properly account for RG evolution effects up to the high scale $M_\phi\sim 1$~TeV. Assuming that either the mixed-chirality or the same-chirality couplings dominate, we derive from the current experimental upper limit $\mbox{Br}(D^0\to\mu^+\mu^-)<7.6\cdot 10^{-9}$ (at 95\% CL) \cite{Aaij:2013cza} the bounds
\begin{equation}\label{Dmumurela}
\begin{aligned}
   \sqrt{\big|\lambda_{c\mu}^L\big|^2 \big|\lambda_{u\mu}^R\big|^2
         + \big|\lambda_{c\mu}^R\big|^2 \big|\lambda_{u\mu}^L\big|^2} 
   &< 1.2\cdot 10^{-3}\,\hat M_\phi^2 \,, \\
   \big| \lambda_{c\mu}^L\lambda_{u\mu}^{L*} + \lambda_{c\mu}^R \lambda_{u\mu}^{R*} \big|
   &< 0.051\,\hat M_\phi^2 \,.
\end{aligned}   
\end{equation}
Compared with \cite{deBoer:2015boa} we obtain a stronger bound on the mixed-chirality couplings, because we include RG evolution effects of the charm-quark mass. On the other hand, a stronger bound (by about a factor~3) than ours on the same-chirality couplings can be derived from the decay $D^+\to\pi^+\mu^+\mu^-$ \cite{deBoer:2015boa,Fajfer:2015mia}. A comprehensive analysis of other rare charm processes along the lines of these references is left for future work. Note that relations (\ref{RDsolu}), (\ref{bound1}) and (\ref{Dmumurela}) can naturally be satisfied assuming hierarchical matrices with $O(1)$ entries for the left-handed couplings and an overall suppression of right-handed couplings. Such a suppression is technically natural, since the right-handed couplings arise from a different operator in the Lagrangian (\ref{Leff}).

\begin{figure}
\includegraphics[width=0.4\textwidth]{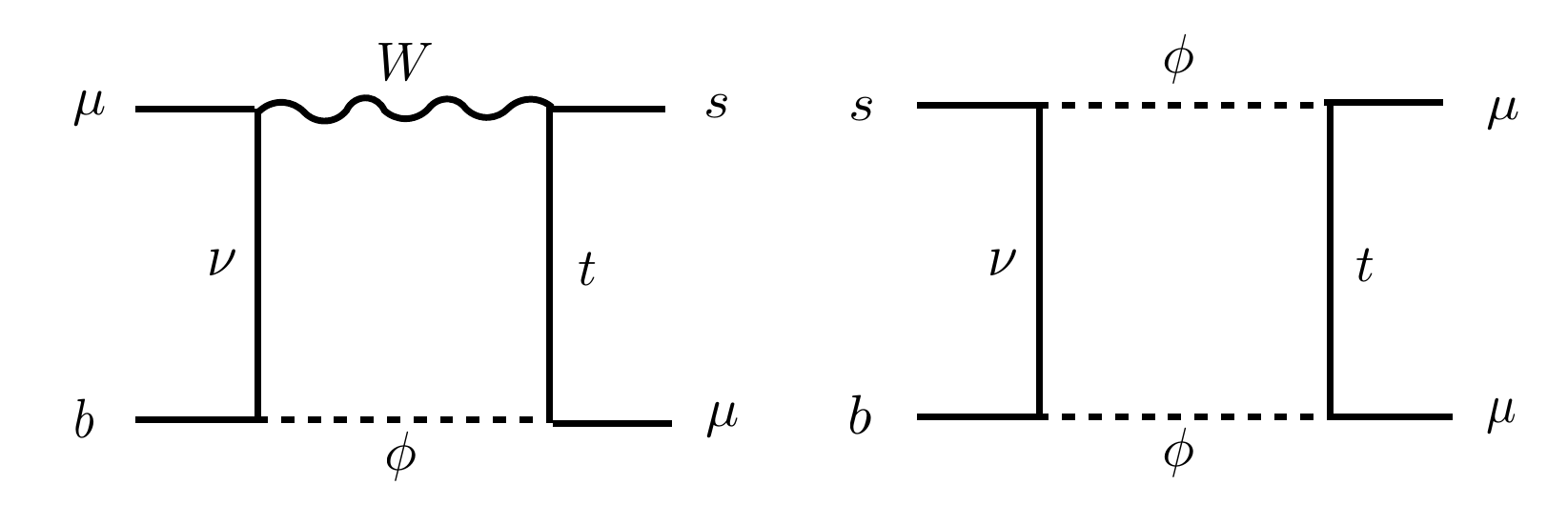}
\vspace{-1mm}
\caption{\label{fig:loops} 
Loop graphs contributing to $b\to s\mu^+\mu^-$ transitions.}
\end{figure}

\vspace{2mm}
{\em Loop-Induced Processes.} 
Earlier this year, LHC$b$ has reported a striking departure from lepton universality in the ratio $R_K$ in (\ref{RK}) \cite{Aaij:2014ora}. Leptoquarks can provide a natural source of flavor universality violation, because their couplings to fermions are not governed by gauge symmetries, see e.g.~\cite{Leurer:1993em,Davidson:1993qk}. A model-independent analysis of this observable was presented in \cite{Hiller:2014yaa,Sahoo:2015wya,Becirevic:2015asa}, while global fits combining the data on $R_K$ with other observables in $b\to s\ell^+\ell^-$ transitions (in particular angular distributions in $\bar B\to\bar K^*\mu^+\mu^-$) were performed in \cite{Hurth:2014vma,Altmannshofer:2014rta,Beaujean:2015gba,Descotes-Genon:2015uva}. The authors of \cite{Hiller:2014yaa,Sahoo:2015wya,Becirevic:2015asa} also studied leptoquark models, in which contributions to $R_K$ arise at tree level. In this case the leptoquark mass is expected to be outside the reach for discovery at the LHC, unless the relevant couplings are very small. In our model effects on $R_K$ arise first at one-loop order from diagrams such as those shown in Figure~\ref{fig:loops}, while we do not find any contributions from flavor-changing $\gamma$ and $Z$ penguins. Working in the limit where $M_\phi^2\gg m_{t,W}^2$, we obtain for the contributions to the relevant Wilson coefficients in the basis of \cite{Hiller:2014yaa} 
\begin{equation}\label{bsmumucoupl}
\begin{aligned}
   C_{LL}^{\mu(\phi)} &= \frac{m_t^2}{8\pi\alpha M_\phi^2} \left| \lambda_{t\mu}^L \right|^2 \\
   &\quad\mbox{}- \frac{1}{64\pi\alpha}\,\frac{\sqrt2}{G_F M_\phi^2}\,
    \frac{\big(\lambda^L\lambda^{L\dagger}\big)_{bs}}{V_{tb} V_{ts}^*}\,
    \big(\lambda^{L\dagger}\lambda^L\big)_{\mu\mu} \,, \\
   C_{LR}^{\mu(\phi)} &= \frac{m_t^2}{16\pi\alpha M_\phi^2} \left| \lambda_{t\mu}^R \right|^2
    \bigg[ \ln\frac{M_\phi^2}{m_t^2} - f(x_t) \bigg] \\
   &\quad\mbox{}- \frac{1}{64\pi\alpha}\,\frac{\sqrt2}{G_F M_\phi^2}\,
    \frac{\big(\lambda^L\lambda^{L\dagger}\big)_{bs}}{V_{tb} V_{ts}^*}\,
    \big(\lambda^{R\dagger}\lambda^R\big)_{\mu\mu} \,,
\end{aligned}   
\end{equation}
where $m_t\equiv m_t(m_t)\approx 162.3$~GeV is the top-quark mass and $f(x_t)=1+\frac{3}{x_t-1}\,\big(\frac{\ln x_t}{x_t-1}-1\big)\approx 0.47$. Analogous expressions hold for $b\to s e^+ e^-$ transitions. The first term in each expression arises from the four mixed $W$--\,$\phi$ box graphs. Relation (\ref{cute}) ensures that the sum of these diagrams is gauge invariant. Importantly, these terms inherit the CKM and GIM suppression factors of the SM box diagrams. The remaining terms result from the box diagram containing two leptoquarks. A good fit to the data can be obtained for $-1.5<C_{LL}^\mu<-0.7$ and $C_{LR}^\mu\approx 0$ at $\mu\sim M_\phi$, assuming that new physics only affects the muon mode -- the  ``one-operator benchmark point'' considered in \cite{Hiller:2014yaa}. In this letter we concentrate on this benchmark point for simplicity. Interestingly, the global fit to all $b\to s\ell^+\ell^-$ data is also much improved for $C_{LL}^\mu\approx -1$ and $C_{LR}^\mu\approx 0$ \cite{Hurth:2014vma,Altmannshofer:2014rta,Beaujean:2015gba,Descotes-Genon:2015uva}, and even the slight deviation in the ratio $\mathrm{Br}(B_s\to\mu^+\mu^-)/\mathrm{Br}(B_s\to\mu^+\mu^-)_\mathrm{SM}=0.79\pm 0.20$ seen in the combination of LHC$b$ \cite{Aaij:2013aka} and CMS \cite{Chatrchyan:2013bka} measurements can be explained. These observations yield further evidence for the suppression of right-handed leptoquark couplings compared with left-handed ones. We will see below that such a pattern is also required by purely leptonic rare processes.

The contributions from mixed $W$--\,$\phi$ box graphs in (\ref{bsmumucoupl}) are controlled by the couplings of the leptoquark to top-quarks and muons. These terms are predicted to be positive in our model and hence alone they cannot explain the $R_K$ anomaly. The contributions from the box graph with two internal leptoquarks are thus essential to reproduce the benchmark value $C_{LL}^\mu\approx -1$. This requires
\begin{equation}\label{rela2}
  \sum_i\big|\lambda_{u_i\mu}^L\big|^2\,
   \mbox{Re}\,\frac{\big(\lambda^L\lambda^{L\dagger}\big)_{bs}}{V_{tb} V_{ts}^*}
   - 1.74\,\big|\lambda_{t\mu}^L\big|^2
  \approx 12.5\,\hat M_\phi^2 \,.
\end{equation}
The analogous combination of right-handed couplings should be smaller, so as to obtain $C_{LR}^\mu\approx 0$. Combining (\ref{rela2}) with the upper bound in (\ref{bound1}) yields
\begin{equation}\label{nicebound}
   \sqrt{\big|\lambda_{u\mu}^L\big|^2 + \big|\lambda_{c\mu}^L\big|^2 
    + \bigg( 1 - \frac{0.77}{\hat M_\phi^2} \bigg) \big|\lambda_{t\mu}^L\big|^2} > 2.36 \,,
\end{equation}
where the top contribution is suppressed for the leptoquark masses we consider. In order to reproduce $C_{LL}^\mu=-0.7$ or $-1.5$ instead of the benchmark value $-1$, the right-hand side of this bound must be replaced by 2.0 or 2.9, respectively. The above condition can naturally be satisfied with a large generation-diagonal coupling $\lambda_{c\mu}^L$.

The ratio $(\lambda^L\lambda^{L\dagger})_{bs}/(V_{tb} V_{ts}^*)$ in (\ref{rela2}) can also be constrained by the existing measurements of the $B_s\!-\!\bar B_s$ mixing amplitude. In our model the new-physics contribution arises from box diagrams containing two leptoquarks, which generate the same operator as in the SM. It is thus useful to follow the suggestion of the UT{\em fit} Collaboration and define the ratio $C_{B_s}\,e^{2i\phi_{B_s}}\equiv\langle B_s|\,H_{\rm eff}^{\rm full}\,|\bar B_s\rangle/\langle B_s|\,H_{\rm eff}^{\rm SM}\,|\bar B_s\rangle$ \cite{Bona:2007vi}. We obtain
\begin{equation}
   C_{B_s}^{(\phi)}\,e^{2i\phi_{B_s}^{(\phi)}} = 1 + \frac{1}{g^4 S_0(x_t)}\,\frac{m_W^2}{M_\phi^2}
    \left[ \frac{\big(\lambda^L\lambda^{L\dagger}\big)_{bs}}{V_{tb} V_{ts}^*} \right]^2 ,
\end{equation}
where $g=\sqrt{4\pi\alpha}/s_W$ is the SU(2) gauge coupling, and $S_0(x_t)=\frac{4x_t-11x_t^2+x_t^3}{4(1-x_t)^2}-\frac{3x_t^3\ln x_t}{2(1-x_t)^3}\approx 2.30$ is the loop function for the SM box diagram. The values obtained from the global fit are $C_{B_s}=1.052\pm 0.084$ and $\phi_{B_s}=(0.72\pm 2.06)^\circ$, which when interpreted as a measurement of leptoquark parameters gives 
\begin{equation}\label{Bsmixval}
   \frac{\big(\lambda^L\lambda^{L\dagger}\big)_{bs}}{V_{tb} V_{ts}^*}
   \approx (1.87+0.45 i)\,\hat M_\phi \,.
\end{equation}
Note that for $M_\phi\lesssim 1$~TeV the central value of the real part of this ratio is close to the upper bound obtained in (\ref{bound1}). At 90\% CL the real part can be as large as $3.6\,\hat M_\phi$, while the phase becomes undetermined. As long as $M_\phi<1.6$~TeV, the upper bound on the real part is thus somewhat weaker than the one obtained from (\ref{bound1}). It is interesting that to reproduce the benchmark value $C_{LL}^\mu\approx -1$ we need a value of $(\lambda^L\lambda^{L\dagger})_{bs}$ close to the upper bound in (\ref{rela2}) and close to the central value in (\ref{Bsmixval}). Our model thus predicts that the $\bar B\to\bar K^{(*)}\nu\bar\nu$ decay rates are enhanced compared with the SM, and that future measurements should find a new-physics contribution to $B_s\!-\!\bar B_s$ mixing close to the current best fit value.

Leptoquark contributions to the dipole coefficient $C_{7\gamma}$ mediating $\bar B\to X_s\gamma$ decays result in 
\begin{equation}
   C_{7\gamma} = C_{7\gamma}^\mathrm{SM} + \bigg( \frac{v}{12M_\phi} \bigg)^2\,
    \frac{\big(\lambda^L\lambda^{L\dagger}\big)_{bs}}{V_{tb} V_{ts}^*} \,.
\end{equation}
Relation (\ref{bound1}) implies that the corresponding change in the $\bar B\to X_s\gamma$ branching ratio is less than about 1\% and thus safely below the experimental bound.

Further constraints on the leptoquark couplings entering (\ref{nicebound}) arise from LEP measurements of the $Z$-boson partial widths into leptons. In particular, we find for the one-loop corrections to the $Z\mu\bar\mu$ couplings 
\begin{align}\label{Zmumu}
   g_A^\mu &=g_A^{\mu,{\rm SM}} 
    \pm \frac{3}{32\pi^2}\,\frac{m_t^2}{M_\phi^2}\,\bigg( \ln\frac{M_\phi^2}{m_t^2} - 1 \bigg)\,
     \big| \lambda_{t\mu}^A \big|^2 \notag \\
   &\quad\mbox{}- \frac{1}{32\pi^2}\,\frac{m_Z^2}{M_\phi^2}\,
    \Big( \big| \lambda_{u\mu}^A \big|^2 + \big| \lambda_{c\mu}^A \big|^2 \Big) \\
   &\quad\mbox{}\times \left[ \bigg( \delta_{AL} - \frac{4 s_W^2}{3} \bigg)\,
    \bigg( \ln\frac{M_\phi^2}{m_Z^2} + i\pi + \frac13 \bigg) - \frac{s_W^2}{9} \right] , \notag
\end{align}
where the upper (lower) sign refers to $A=L$ ($R$). For simplicity we have set $m_Z^2/(4m_t^2)\to 0$ in the top contribution, which numerically is a good approximation. We require that the $Z\to\mu^+\mu^-$ partial width agrees with its SM value within $2\sigma$ of its experimental error. Assuming that the left-handed couplings are larger than the right-handed ones, and that a single coupling combination dominates, we obtain
\begin{equation}\label{Zbounds}
   \sqrt{\big|\lambda_{c\mu}^L\big|^2 + \big|\lambda_{u\mu}^L\big|^2} 
    < \frac{3.24\,\hat M_\phi}{b_{cu}^{1/2}} \,, \quad
   \big|\lambda_{t\mu}^L\big| < \frac{1.22\,\hat M_\phi}{b_t^{1/2}} \,,
\end{equation}
where $b_{cu}=1+0.39\,\ln\hat M_\phi$ and $b_t=1+0.76\,\ln\hat M_\phi$.
The first relation is compatible with the bound (\ref{nicebound}) as long as $M_\phi>0.67$~TeV.

\begin{figure}
\includegraphics[width=0.4\textwidth]{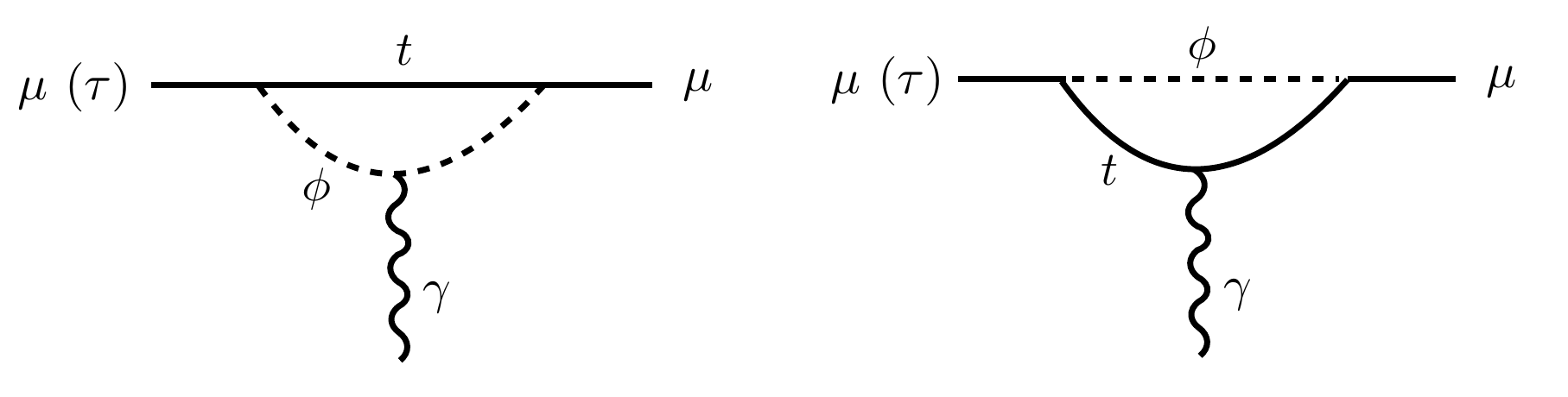}
\vspace{-1mm}
\caption{\label{fig:ref}
Loop diagrams contributing to $(g-2)_\mu$ and $\tau\to\mu\gamma$.}
\end{figure}

\begin{figure*}[t]
\begin{center}
\begin{tabular}{ccc}
\includegraphics[width=0.38\textwidth]{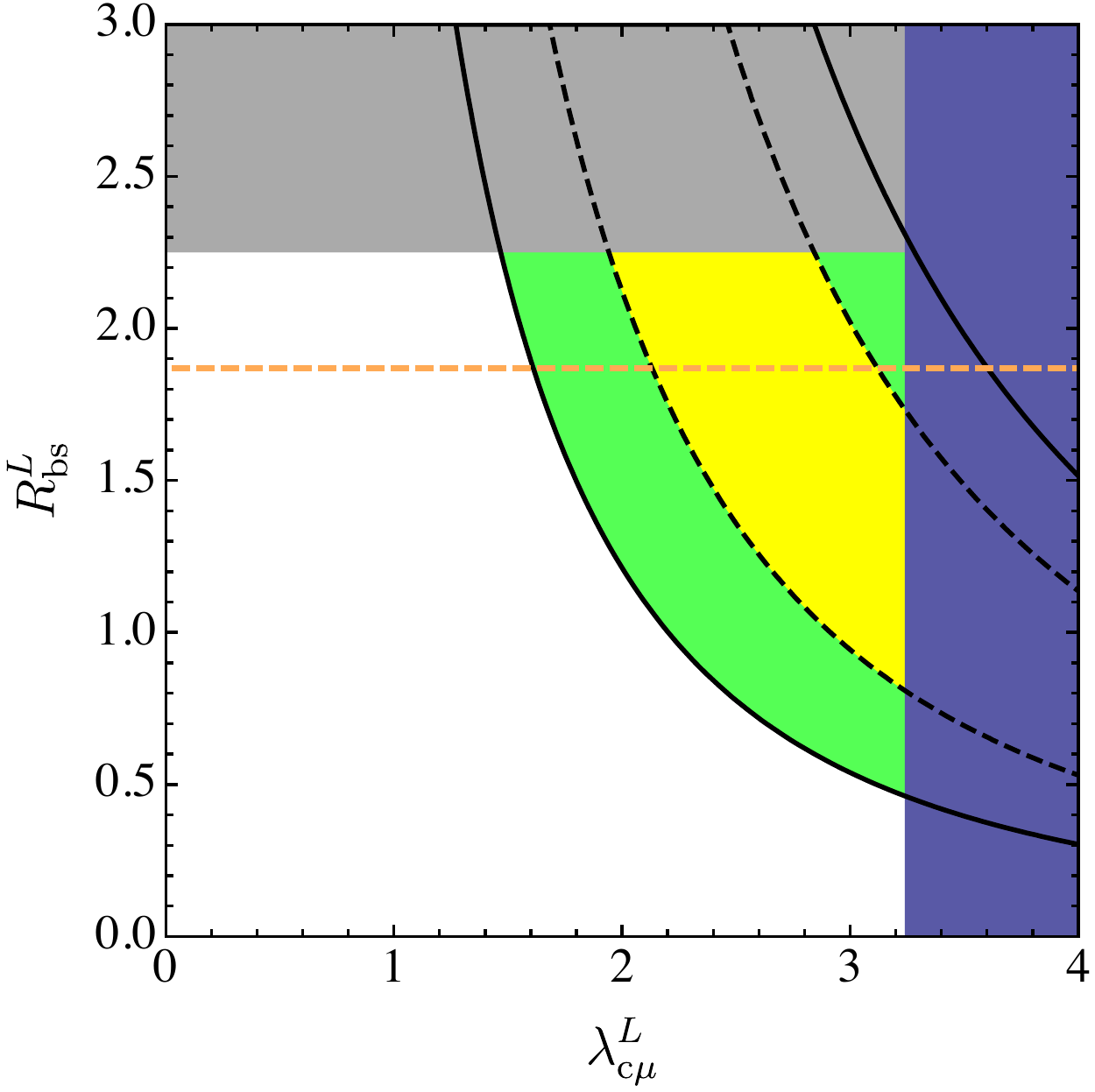} & ~~~~~ &
\includegraphics[width=0.38\textwidth]{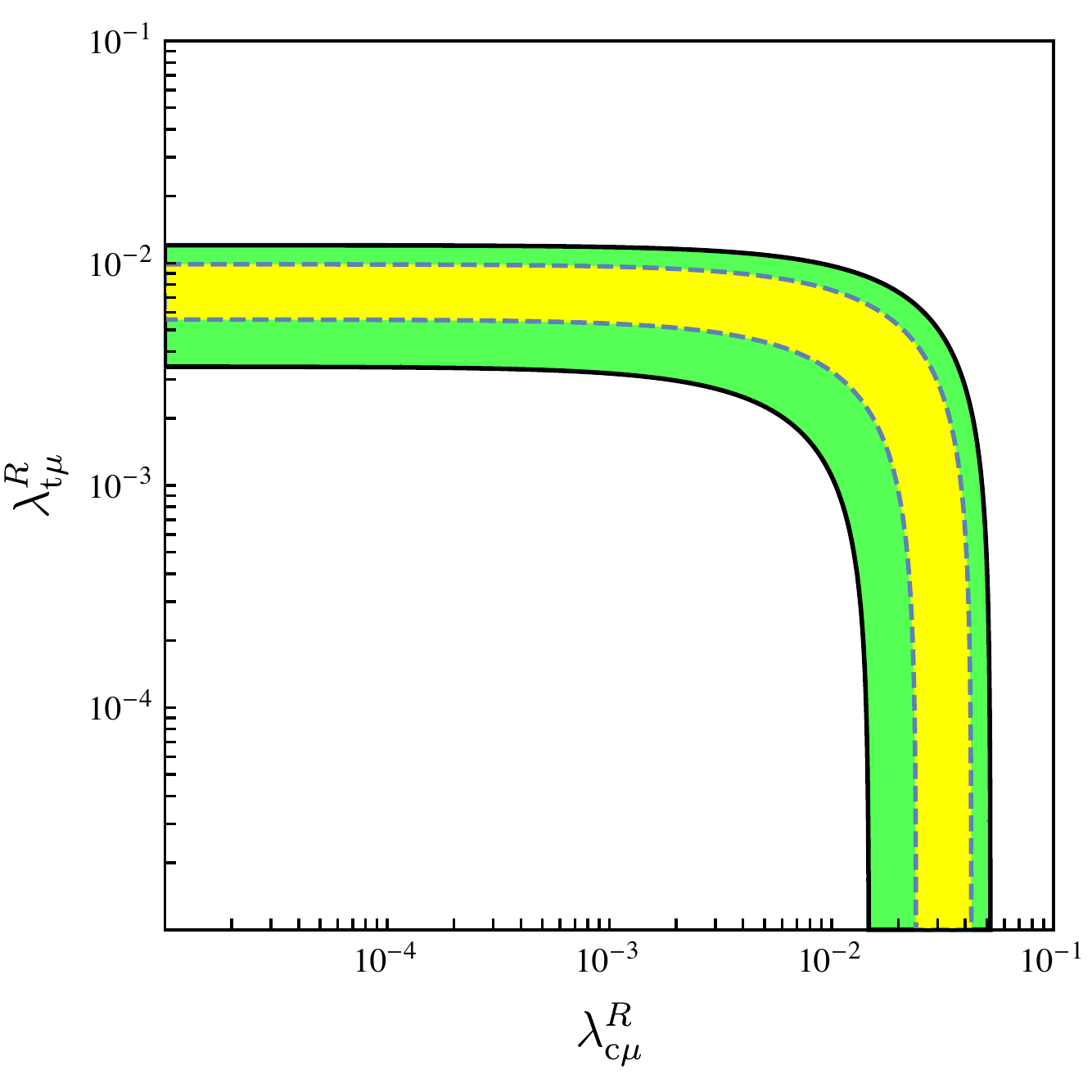}
\end{tabular}
\caption{\label{fig:suppl} 
Fits to $R_K$ (left panel) and $(g-2)_\mu$ (right panel). The $1\sigma$ ($2\sigma$) fit region is shown in yellow (green), and excluded regions from $R_{\bar\nu\nu}$ and $Z\to\mu^+\mu^-$ are shaded gray and blue, respectively.}
\end{center}
\end{figure*}

The couplings of the muon to up-type quarks, which enter in (\ref{bsmumucoupl}), also contribute to the muon anomalous magnetic moment $a_\mu=(g-2)_\mu/2$ and the rare decay $\tau\to\mu\gamma$. In our model, new-physics contributions to these quantities arise from the one-loop vertex corrections shown in Figure~\ref{fig:ref}. Working in the limit where $M_\phi^2\gg m_t^2$, we obtain in agreement with \cite{Djouadi:1989md,Chakraverty:2001yg,Cheung:2001ip} 
\begin{equation}\label{eq:amu}
\begin{aligned}
   a_\mu^{(\phi)} &= \sum_{q=t,c} \frac{m_\mu m_q}{4\pi^2 M_\phi^2}\,
    \bigg( \ln\frac{M_\phi^2}{m_q^2} - \frac74 \bigg)\, 
    \mbox{Re}\big( \lambda_{q\mu}^R\lambda_{q\mu}^{L*} \big) \\
   &\quad\mbox{}- \frac{m_\mu^2}{32\pi^2 M_\phi^2} \left[ \big(\lambda^{L\dagger}\lambda^L\big)_{\mu\mu}
    + \big(\lambda^{R\dagger}\lambda^R\big)_{\mu\mu} \right] ,
\end{aligned}
\end{equation}
where $m_q\equiv m_q(m_q)$ are running quark masses. The present experimental value of $a_\mu$ differs from the SM prediction by $(287\pm 80)\cdot 10^{-11}$ \cite{Davier:2010nc}. The last term above is negative and thus of wrong sign, however it is suppressed by the small muon mass. Assuming the worst case, where the first bound in (\ref{Zbounds}) is saturated, this term contributes approximately $-37\cdot 10^{-11}$. To reproduce the observed value in our model, we must then require that (we use $m_c\approx 1.275$~GeV)
\begin{equation}\label{amubound}
   a_c\,\mbox{Re}\big( \lambda_{c\mu}^R\lambda_{c\mu}^{L*} \big) 
   + 20.7 a_t\,\mbox{Re}\big( \lambda_{t\mu}^R\lambda_{t\mu}^{L*} \big)
   \approx 0.08\,\hat M_\phi^2 \,,
\end{equation}
where $a_t=1+1.06\,\ln\hat M_\phi$ and $a_c=1+0.17\,\ln\hat M_\phi$. Assuming hierarchical coupling matrices and a suppression of right-handed couplings compared with left-handed ones, as mentioned earlier, both terms on the left-handed side can naturally be made of the right magnitude to explain the anomaly. We stress that $a_\mu$ is the only observable studied in this letter which requires a non-zero right-handed coupling of the leptoquark. For example, if (\ref{nicebound}) is satisfied with $|\lambda_{c\mu}^L|\sim 2.4$, the $a_\mu$ anomaly can be explained with $|\lambda_{c\mu}^R|\sim 0.03$. The leptoquark contribution to $a_\mu$ is tightly correlated with one-loop radiative corrections to the masses of the charged leptons. Relation (\ref{amubound}) ensures that these corrections stay well inside the perturbative regime. The Wilson coefficients of the dipole operators mediating the radiative decay $\tau\to\mu\gamma$ are given by expressions very closely resembling those in (\ref{eq:amu}) \cite{Chakraverty:2001yg,Cheung:2001sb}. From the current experimental bound $\mbox{Br}(\tau\to\mu\gamma)<4.4\cdot 10^{-8}$ at 90\% CL \cite{Aubert:2009ag}, we obtain
\begin{equation}\label{taubound}
\begin{aligned}
   &\bigg[ \Big| a_c\,\lambda_{c\tau}^R\lambda_{c\mu}^{L*} + 20.7 a_t\,\lambda_{t\tau}^R\lambda_{t\mu}^{L*} 
    - 0.015 \big(\lambda^{L\dagger}\lambda^L\big)_{\mu\tau} \Big|^2 \\[-1mm]
   &\hspace{2mm}+ (L\leftrightarrow R) \bigg]^{1/2} \!< 0.017\,\hat M_\phi^2 \,.
\end{aligned}
\end{equation}
The mixed-chirality contributions are naturally very small, because they each involve one off-diagonal and one right-handed coupling. The even-chirality contributions involve at least one off-diagonal coupling, which makes them small enough to satisfy the bound. Barring a fine-tuning, relation (\ref{taubound}) implies that $\big|\lambda_{t\tau}^R\lambda_{t\mu}^{L*}\big|^2+\big|\lambda_{t\tau}^L\lambda_{t\mu}^{R*}\big|^2<6\cdot 10^{-7}$ (for $M_\phi\sim 1$~TeV). Assuming that this value is saturated, we obtain a $h\to\mu^\pm\tau^\mp$ branching fraction ranging between $10^{-9}$ and $10^{-7}$ for $g_{h\phi}$ of $O(1)$ and $O(4\pi)$, respectively, which is many orders of magnitude smaller than the central value of 0.84\% reported by CMS \cite{Khachatryan:2015kon}. This finding is in accordance with a model-independent argument made in \cite{Dorsner:2015mja,Altmannshofer:2015esa}. It is possible to evade this conclusion by means of excessive fine tuning, e.g.\ by precisely tuning the three contributions on the left-hand side of (\ref{taubound}) to cancel each other, or by engineering analogous cancellations by introducing a second leptoquark \cite{Dorsner:2015mja,Cheung:2015yga}.

\vspace{2mm}
{\em Conclusions.} 
We have argued that the violation of lepton universality observed in $R_K$ and $R_{D^{(*)}}$ can be explained by extending the SM with a single scalar leptoquark with mass $M_\phi\sim 1$~TeV and $\mathcal{O}(1)$ generation-diagonal couplings to SU$(2)_L$ doublet quarks and leptons. FCNC constraints from $B_s-\bar B_s$ mixing, $D\to\mu^+\mu^-$ and $\tau\to\mu\gamma$ result in upper bounds of order $10^{-1}\!-\!10^{-2}$ on the corresponding generation off-diagonal couplings, and of order $10^{-2}\!-\!10^{-3}$ on the couplings to SU$(2)_L$ singlet quarks and leptons. Remarkably, a coupling $|\lambda_{c\mu}^R|\sim 0.03$ explains the anomalous value of $a_\mu$ without introducing further constraints. A graphical illustration of the allowed parameter space for the most relevant leptoquark couplings of our model is shown in Figure~\ref{fig:suppl}. The plot on the left shows the parameter space preferred by a fit to $R_K$ for the couplings $R_{bs}^L\equiv\text{Re}[(\lambda^L\lambda^{L\dagger})_\mathrm{bs}/(V_{tb} V_{ts}^\dagger)]$ and $\lambda^L_{c\mu}$ and a benchmark mass of $M_\phi=1$~TeV in yellow and green, and the bounds from corrections to the $Z\to\mu^+\mu^-$ couplings and $R_{\nu\bar \nu}$ in blue and gray, respectively. The best fit to the $B_s\!-\!\bar B_s$ mixing amplitude is shown by the dashed orange line. The right plot shows a fit to $(g-2)_\mu$ in the $\lambda^R_{c\mu}\!-\!\lambda^R_{t\mu}$ plane for $M_\phi=1$~TeV, $\lambda^L_{c\mu}=2.4$ and $\lambda^L_{t\mu}=0.5$. For the small right-handed couplings necessary to explain the $(g-2)_\mu$ anomaly, no constraints from any other observables arise. The couplings entering $R_{D^{(*)}}$ are independent from the ones relevant for either $R_K$ and $(g-2)_\mu$. 

Our model makes several characteristic predictions, including a correction to $\mbox{Br}(Z\to\mu^+\mu^-)$ at least at the current $1\sigma$ level, as well as sizeable and correlated effects in $B_s\!-\!\bar B_s$ mixing and $\bar B\to\bar K^{(*)}\nu\bar\nu$ decays. If the anomalies persist, this pattern of deviations results in a complementary discovery potential at Run-II of the LHC, the upcoming Belle-II experiment and a future FCC-ee (TLEP) collider.

\vspace{3.3mm}
{\em Acknowledgements.} 
M.B.~acknowledges support of the Alexander von Humboldt Foundation. M.N.~is supported by the ERC Advanced Grant EFT4LHC, the PRISMA Cluster of Excellence EXC1098 of DFG, and grant 05H12UME of BMBF. We thank Sophie Renner for a useful conversation and Xin Zhang for pointing out a typo in (\ref{eq:sign}).

\end{document}